\documentclass[12pt]{article}
\usepackage{amsmath,amssymb,graphicx,color}

\newwrite\ffile\global\newcount\figno \global\figno=1

\def\writedef#1{}

\input epsf
\def\figin{\epsfcheck\figin}\def\figins{\epsfcheck\figins}
\def\epsfcheck{\ifx\epsfbox\UnDeFiNeD
\message{(NO epsf.tex, FIGURES WILL BE IGNORED)}
\gdef\figin##1{\vskip2in}\gdef\figins##1{\hskip.5in}
\else\message{(FIGURES WILL BE INCLUDED)}%
\gdef\figin##1{##1}\gdef\figins##1{##1}\fi}

\def\figinsert{}
\def\ifig#1#2#3{\xdef#1{fig.~\the\figno}
\writedef{#1\leftbracket fig.\noexpand~\the\figno}%
\figinsert\figin{\centerline{#3}}\medskip\centerline{\vbox{\baselineskip12pt
\advance\hsize by -1truein\center\footnotesize{  Fig.~\the\figno.}
#2}}
\bigskip\endinsert\global\advance\figno by1}
\def\endinsert{}

\providecommand{\beq}{}
\renewcommand{\beq}{\begin{equation}}
\providecommand{\eeq}{}
\renewcommand{\eeq}{\end{equation}}

\usepackage{amssymb}
\usepackage{graphics}
\usepackage{graphicx}

\begin{document}

\begin{center}
\vskip 2cm \Large {\bf AdS/QCD Phenomenological Models from a
Back-Reacted Geometry} \vskip 1.5cm \large Jonathan P. Shock
\footnote[1]{\noindent \tt
 jps@itp.ac.cn}, Feng
Wu \footnote[2]{\noindent \tt
 fengwu@itp.ac.cn}, Yue-Liang Wu \footnote[3]{\noindent \tt
 ylwu@itp.ac.cn}\small and \large Zhi-Feng Xie \footnote[4]{\noindent \tt
 xzhf@itp.ac.cn}\vskip 1cm \small
{\em Kavli Institute for Theoretical Physics China (KITPC) \\
Institute of Theoretical Physics, Chinese Academy of Sciences\\
P.O. Box 2735, Beijing 100080, China}

\end{center}
\vskip 2cm
\begin{abstract}
We construct a fully back-reacted holographic dual of a
four-dimensional field theory which exhibits chiral symmetry
breaking. Two possible models are considered by studying the effects
of a five-dimensional field, dual to the $q\bar{q}$ operator. One
model has smooth geometry at all radii and the other dynamically
generates a cutoff at finite radius. Both of these models satisfy
Einstein's field equations. The second model has only three free
parameters, as in QCD, and we show that this gives
phenomenologically consistent results. We also discuss the
possibility that in order to obtain linear confinement from a
back-reacted model it may be necessary to consider the condensate of
a dimension two operator.
\end{abstract}
\thispagestyle{empty}

\newpage
\pagenumbering{arabic}

\section{Introduction}

Strong interactions are described in the standard model (SM) by an
$SU(3)$ gauge theory known as quantum chromodynamics (QCD)
\cite{Fritzsch:1973pi}. Quarks, which are basic constituents in the
SM, are in the fundamental representation of the $SU(3)$ group. They
interact with each other by coupling to the gluons, the vector gauge
bosons transforming as the adjoint representation of $SU(3)$. Since
the gauge group is non-Abelian the gluons have direct
self-interactions (this feature is closely related to the origin of
life). It is these self-interactions that cause a negative beta
function, $\beta(\mu)$, for the running coupling constant
$\alpha_{s}(\mu)$ \cite{Gross:1973id,Politzer:1973fx}. This is a
unique feature among four-dimensional renormalizable gauge theories.
Because $\beta(\mu) < 0$, the coupling constant $\alpha_{s}(\mu)$
decreases at short distances (UV region) and it is this
anti-screening effect, known as asymptotic freedom, which makes the
UV region perturbatively understood.

At low energies (IR region), solving the theory becomes extremely
difficult. As the coupling constant $\alpha_{s}(\mu)$ grows in the
IR, perturbative methods are no longer applicable. We are currently
unable to solve from first principle the low energy dynamics of QCD.
One can only study the low energy features of strong interactions
from experimental data and construct effective quantum field
theories to characterize the low energy features of QCD, such as
dynamically generated spontaneous symmetry breaking
\cite{Nambu:1960xd}. It was recently shown in ref. \cite{DW} that
the gap equation can be derived as the minimal condition of the
effective Higgs potential generated dynamically and the lightest
scalar mesons as the parters of pseudoscalar mesons play the role of
the composite Higgs bosons to realize the dynamical spontaneous
symmetry breaking. As a consequence, the mass spectrum of scalar and
pseudoscalar mesons and their mixings  as well as the relevant low
energy parameters were predicted \cite{DW} to be consistent with the
experimental data by using the same number of input parameters as in
QCD. At low energies, quarks are bound to form hadrons, which are
color singlets composed of quarks and gluons. Isolated quarks and
gluons appear not to exist. This so called color confinement
phenomenon has been a well known conjecture since the development of
QCD and has not yet been proved.

The details of hadronization are still unclear. In the massless
limit, the theory has only one parameter, the coupling constant. If
QCD is the fundamental theory of the strong interactions, it must
have an intrinsic scale to account for the short-range feature of
the strong interactions, even if all the fields are massless. This
is a pure quantum effect and has the consequence that the lightest
hadron is massive. In the QCD Lagrangian, one can see that quark
masses act as sources of the operators $\bar{q}(x)q(x)$. This means
that if operators $\bar{q}(x)q(x)$ are allowed, quarks must be
massive. In other words, $\bar{q}(x)q(x)$ must break some symmetry
in the massless limit. It is not difficult to see that this symmetry
in the massless limit is the global $SU(N_f)_{L} \bigotimes
SU(N_f)_{R}$ chiral symmetry. Therefore, it is natural to expect
that the formation of the condensate $\langle \bar{q}(x) q(x)
\rangle$ will dynamically break the chiral symmetry. None of the
non-perturbative features mentioned above has yet been justified
from first principles.

In \cite{'tHooft:1973jz} 't Hooft proposed the idea that one might
be able to solve QCD by first studying an $SU(N)$ gauge theory in
the limit $1/N \rightarrow 0$. One can then perform a $1/N$
expansion and study the theory perturbatively. In this limit, one
can argue that properties such as color confinement and dynamical
chiral symmetry breaking still exist. Furthermore, the spectra are
given by an infinite number of meson and glueball resonances. Mesons
are stable and the decay amplitudes are $O(1/ \sqrt{N})$. The large
$N$ theory can be treated as an effective theory of hadrons and
glueballs. Since the intrinsic scale $\Lambda_{QCD}$ is independent
of $N$ when $1/N \rightarrow 0$, the mass of mesons does not change
with $N$. It can be shown that current-current correlation functions
can be written as an infinite sum over meson resonances. The poles
and the corresponding residues give the mass and decay constant of
mesons, if one knows how to calculate the two-point function.

The duality between gravity and gauge theories conjectured by
Maldacena \cite{Maldacena:1997re} and further developed in
\cite{Gubser:1998bc,Witten:1998qj} has shed new light on the problem
of strongly coupled gauge theories. Although the original conjecture
in \cite{Maldacena:1997re} is the mathematical correspondence
between the low energy approximation of type ${\rm II}$B string
theory on $AdS_5 \bigotimes S^{5}$ and ${\mathcal{N}} = 4$ $U(N)$
SYM for large $N$ in four dimensions, the holographic principle can
in theory be applied to general gauge theories and their gravity
duals. The idea goes as follows. Consider a manifold $\mathcal{M}$
whose limit in infinity is $\mathcal{X} \bigotimes \mathcal{W}$ with
$\mathcal{X}$ being an Einstein manifold and $\mathcal{W}$ being
compact. After compactification on $\mathcal{X}$, one can build a
manifold $\bar{\mathcal{X}}$ with boundary $\mathcal{S}$ such that
the metric on $\mathcal{X}$ has a double pole near the boundary
$\mathcal{S}$. Then the isometry group of $\mathcal{X}$ acts as the
conformal group of $\mathcal{S}$. The duality is the holographic
correspondence that for a conformal field theory (CFT) on
$\mathcal{S}$, one can find a gravity theory on $\mathcal{X}$ such
that the generating functional of the connected Greens functions in
CFT on $\mathcal{S}$ is equivalent to the action of the gravity
theory on $\bar{\mathcal{X}}$, whose fields on $\mathcal{S}$ are
identified to be the sources of the operators in CFT (see
\cite{hep-th/0608089} for recent work exploring generalisations of
this relationship). The mass $m_{\phi}$ of a $p$-form field $\phi$
on $\bar{\mathcal{X}}$ is quantized and related to the conformal
dimension $\Delta$ of the operator $\hat{O}_{\phi}$ on $\mathcal{S}$
it corresponds to. The relation is \cite{Witten:1998qj}
$m_{\phi}^{2}=(\Delta-p)(\Delta+p-d)$ where $d$ is the dimension of
$\mathcal{S}$. The anomalous dimension of the operator
$\hat{O}_{\phi}$ on $\mathcal{S}$ corresponds to the quantum
corrections on $m_{\phi}$. That is, the quantum corrections on
$m_{\phi}$, $\delta m_{\phi}$, should satisfy
\[
\delta m_{\phi}=( \Delta_{0} - \frac{d}{2} )
\frac{\gamma_{\hat{O}_{\phi_{0}}} }{m_{\phi_{0}}}
\]
where $\gamma_{\hat{O}_{\phi_{0}}}$ is the anomalous dimension of
$\hat{O}_{\phi_{0}}$ and we use the subscript $0$ to denote the
unrenormalized quantities. If the anomalous dimension of an operator
vanishes, the operator must be a generator of some symmetry algebra
on $\mathcal{S}$. In this case, the $m_{\phi}$ must be protected
from corrections by some symmetry on $\bar{\mathcal{X}}$. For
example, the dimension 3 operators $\bar{q}\gamma^{\mu} t^{a} q$ are
conserved currents in QCD. Their holographic dual are 1-form fields
$A_{\mu}^{a}$ whose masses vanish from the relation mentioned above.
$A_{\mu}^{a}$ must be vector gauge fields on $\bar{\mathcal{X}}$ and
remain massless by gauge symmetry.

QCD is clearly not a conformal field theory. However, due to the
asymptotic freedom, it approaches to the conformal limit in UV
regions. At high energies, one may neglect quark masses and the
chiral symmetry is restored. In these regions, when approaching the
fixed point, which is suggested to exist
\cite{hep-ph/9705242,hep-th/0412330}, QCD becomes a conformal field
theory and the holographic recipe is applicable. Many efforts have
been made in constructing a holographic dual of QCD. See
\cite{hep-ph/0501128,hep-ph/0501218,hep-ph/0507049,hep-ph/0510240,hep-ph/0510268,hep-ph/0510334,hep-ph/0512089,hep-ph/0512240,hep-ph/0602177,hep-ph/0602229,hep-ph/0603249,
hep-ph/0603142,hep-ph/0608266,hep-ph/0609112} and references therein
for various approaches. Usually in these models, the background
manifold is chosen to be $AdS$, which means one did not consider the
effects of the condensates. This, however, is not logically
consistent. In QCD, the relevant and marginal operators that can
form condensates without breaking Lorentz and gauge symmetries at
low energies are $F_{\mu\nu}^{a} F^{a \mu\nu}$ (dimension 4) and
$\bar{q} q$ (dimension 3). The dimension 2 operator condensate
$\langle A_{\mu}^{a} A^{a \mu} \rangle_{min} $\cite{Gubarev:2000eu}
will be discussed later. These condensates can be used to
parameterize the non-perturbative observables and their effects
should not be neglected. The holographic dual of these operators are
scalar fields, acting as sources for these operators on the boundary
of $\bar{\mathcal{X}}$. Including the effects of these condensates
means we are introducing these scalar fields in the bulk, which in
turn will deform the $AdS$ manifold. In this article, we consider
the effects of these scalar fields and solve the graviton-tachyon
system to give a correct deformed $AdS$ metric and perform a
consistent analysis. Adopting this bottom-up approach, one is
running a risk that the model built might not have anything to do
with the underlying theory. Nevertheless, if the model can describe
the distinctive features of QCD and accurately predict the
non-perturbative quantities such as mass spectrum and decay
constants of hadrons, then it deserves one's attention and might be
able to give us some clues about the structure of the underlying
theory if one believes that QCD has a holographic dual theory.

\section{Effects of The Condensates}

Before constructing a phenomenological model, we will first probe
the general effects of a bulk field $\phi(x^{\mu},y)$ on the metric,
at the same time setting up the notation. We start with the
following five-dimensional effective action
\begin{equation}
\label{action} S=\int_{\bar{\mathcal{X}}} d^5 x \sqrt{g} ( -R +
\frac{1}{2} (\partial \phi )^2 +V(\phi) )\ ,
\end{equation}
where $R$ is the five-dimensional Ricci scalar.

With time-reversal and parity symmetries, one can show that the
static metric which respects Poincaré symmetry on four-dimensional
coordinates $x^{\mu}$, the one we are interested, has the following
form
\begin{equation}
ds^2= e^{-2A(y)} d x_{\mu} d x^{\mu} -dy^2\ .
\end{equation}
As we stressed before, we are building a holographic model based on
our knowledge of QCD, the potential $V(\phi)$ is unknown and will be
determined once the specific form of $\phi(x^{\mu}, y)$ is chosen to
describe some QCD condensate. Without matter fields, $\mathcal{X}$
is an Einstein manifold. Demanding $\mathcal{X}$ to be $AdS_{5}$ in
that case, it is easy to show that the warp factor $A(y)=y$ and the
constant potential $V$ equals $\Lambda=12$ in our units. (With
arbitrary dimension $D$ of $\mathcal{X}$, $\Lambda=(D-1)(D-2)$.)

The equations of motion for Eqn.(\ref{action}) are
\begin{equation}
{1\over 2} g_{PQ} [-R+\frac{1}{2} \partial_{M} \phi \partial^{M}
\phi + V(\phi)] +R_{PQ} -\frac{1}{2} \partial_{P} \phi \partial_{Q}
\phi =0\ ,
\end{equation}
and
\begin{equation}
{\partial V(\phi) \over \partial \phi} = {1 \over \sqrt{g}}
\partial_{P} ( \sqrt{g} g^{PQ} \partial_{Q} \phi )\ ,
\end{equation}
which gives
\begin{equation}
\label{5} 6 A^{''}(y) -12 A^{'2}(y) + V(\phi) -{1\over 2} \phi^{'2}
=0\ ,
\end{equation}
\begin{equation}
\label{6}
 12 A^{'2}(y) - V(\phi) ={1\over 2} \phi^{'2}\ ,
\end{equation}
\begin{equation}
\label{7}
 \phi^{''}(y) - 4A^{'}(y)\phi^{'}(y)+{\partial V(\phi) \over \partial \phi} = 0\ .
\end{equation}
Here we consider the particular solution such that $\phi$ is
independent of $x^{\mu}$, since we demand QCD condensates are
translationally invariant. The equations can be solved using the
method in \cite{Freedman:1999gp}. Eqn.(\ref{5}) and Eqn.(\ref{6})
gives
\begin{equation}
\label{eq.8} 6A^{''}-\phi^{'2}=0\ ,
\end{equation}
and
\begin{equation}
\label{9} 3A^{''}-12 A^{'2} + V(\phi)=0\ .
\end{equation}
Note that care must be taken here as the factor of $6$ in
Eqn.(\ref{eq.8}) becomes a $3$ in the case of a complex field which
is used in section \ref{sec.model}. This factor also affects the
following equations in this section. As we know that without the
scalar field $\phi$, $V=\Lambda$ and $A^{'}(y)=1$ (maximally
symmetric). Therefore, it is natural to expect that
$A^{'}(y)=W(\phi(y))$ after the effect of $\phi$ is considered.
Eqn.(\ref{eq.8}) tells us that $6 {\partial W \over
\partial \phi} = \phi^{'}$ since $A^{''}={\partial W \over \partial
\phi} \phi^{'} $. Eqn.(\ref{9}) can be taken as the definition of
$V(\phi)$, that it, $V(\phi)= 12 W^2-18 ({\partial W \over \partial
\phi})^2$. It is trivial to check that Eqn.(\ref{7}) will
automatically be satisfied:
\[
{\partial V(\phi) \over \partial \phi}= 24 W(\phi) {\partial W(\phi)
\over \partial \phi}-36 {\partial W(\phi) \over \partial \phi}
{\partial^2 W (\phi)\over \partial \phi^2} =4A^{'}\phi^{'}
-\phi^{''}\ .
\]
In the UV region ($y\rightarrow - \infty$), the effects of
condensates are less and less important so we should expect $
\phi(y) \rightarrow 0$ as $y \rightarrow - \infty$. In fact, from
Eqn.(\ref{7}) one can see that when $y \rightarrow - \infty $,
keeping only the first nontrivial term of $V(\phi)$ (i.e., up to
$O(\phi^2)$), the linearized equation of motion for $\phi(y)$ gives
$\phi(y) \sim e^{ny}$. For $\phi(y)=a e^{ny}$, $\phi^{'}=n \phi$.
And it is easy to show that $W(\phi)={n \over 12}\phi^2 + 1$ so that
$A(y)= {a^2 \over 24} e^{2ny} + y $. Therefore, we arrive at an
effective action whose potential $V(\phi)$ equals $V(\phi)=12+
(2n-{n^2 \over 2}) \phi^2 +{n^2 \over 12} \phi^4$. This can be
easily generalized to the case with $n$ scalar fields $\phi_{i}$.
For $\phi_{i}=a_{i} e^{n_{i} y}$, the potential $V(\phi_{i})$ is
\[
V(\phi_{i})=12 + \sum_{i} ( 2 n_{i}-{1 \over 2} n_{i}^{2} )
\phi_{i}^{2} + {1 \over 6} \sum_{i < j} n_{i} n_{j} \phi_{i}^{2}
\phi_{j}^{2} +{1 \over 12} \sum_{i} n_{i}^{2} \phi_{i}^{4},
\]
and the warp factor $A(y)$ is
\[
 A(y)=y+{1 \over 24} \sum_{i} a_{i}^{2} e^{2 n_{i} y}.
 \]
The constant term guarantees that without $\phi_{i}$, $\mathcal{X}$
goes back to $AdS_{5}$. The mass terms only depend on $n_{i}$. The
model we build in this way should be the holographic dual of a
four-dimensional theory. Note that this example for $\phi$ is an
oversimplification and a true dual field will include at least two
terms. Each field $\phi_{i}$ on the UV boundary should act as the
source of some dimension $\Delta_{i}$ operator on the
four-dimensional theory. According to \cite{Witten:1998qj},
$m_{\phi_{i}}^{2}=\Delta_{i} (\Delta_{i}-4)$. Therefore, we have
$n_{i}=\Delta_{i}$ or $4-\Delta_{i}$ and the meaning of $n_{i}$ is
clear now. Also from the mass terms one can see that the condensates
formed from relevant operators correspond to tachyonic scalar fields
on the holographic theory. As long as one demands (i) the warp
factor $A(y)=y$ (i.e., $AdS_{5}$) when matter fields $\phi_{i}(y)=0$
and (ii) $\phi_{i}(y) \sim a_{i} e^{\Delta_{i} y}$ (or
$e^{(4-\Delta_{i}) y}$) in the UV ($y\rightarrow - \infty$) in order
to do the correspondence, the first two terms of the potential
$V(\phi_{i})$ are determined, independent of the specific forms of
$\phi_{i}$. From this example one might think that the coupling
constants of the self-interaction terms only depend on $n_{i}$.
This, however, is in general not true.

In the following section, we will use the method discussed in this
section to construct a phenomenological model for mesons.

\section{The Phenomenological Model}\label{sec.model}

We wish now to consider the phenomenological implications of
including a back-reaction of a five-dimensional field on the
originally AdS geometry. We must first decide which fields are
relevant to the theory in question. In QCD there are four relevant
operators which we may consider given by
$\bar{q}^\alpha_Lq^\beta_R$, $\bar{q}_{L,R}\gamma^\mu t^a q_{L,R}$,
$F^2$ and $\tilde{F}F$. These have been discussed in the literature
in \cite{hep-ph/0501128,hep-ph/0501218}. In this paper we
concentrate on the complex scalar, dimension three operator to study
its effect on the geometry and subsequently the phenomenology of the
model. The inclusion of the source for a dimension four operator in
addition is no more complicated but it does introduce another free
parameter, the condensate of $F^2$. Though this operator is clearly
important, we can argue that its effect will be of a similar order
of magnitude to that of the dimension three operator.

We define the source for the $\bar{q}_L^\alpha q_R^\beta$ operator
as a matrix valued field $X(x_\mu,y)$ proportional to the unit two
by two matrix. This can be extended to three flavours but for now we
discuss only two. The action is given by:
\begin{eqnarray}
S&=&\int_{\bar{\chi}}d^5x\sqrt{g}Tr\left[-R+|DX|^2-V(X)-\frac{1}{4g_5^2}(F_L^2+F_R^2)\right]\\
V(X)&=&-3|X|^2+V_{int}(X)
\end{eqnarray}
where the self-interaction terms are to be determined by a
consistent back-reaction and the vector and axial vector gauge
fields are sources of the vector and axial vector currents in the
four-dimensional theory.

We must determine a consistent form for $X$ which will provide a
sound phenomenological model. There are several distinct routes one
can take at this point, each of which has different repercussions
for the model. The first option is to make an ansatz for $X$ which
is polynomial in $z(=e^y)$ with a finite number of terms. The second
option is to construct a function of $z$ which has an infinite
series expansion but has no poles. The third option is to define a
function which is singular at some specific point $z$ in the
geometry. Each of these will be discussed in turn.

We know that the AdS behavior of the field $X$, corresponding to a
dimension three scalar operator must go like:
\begin{equation}\label{eq.X}
X=\left(\frac{m_q}{2}z+\frac{\sigma}{2}z^3\right){\mathbb
I}_{2\times 2} ,
\end{equation}
where the coefficients can be shown to correspond to a quark mass
and bilinear condensate respectively. If we wish to construct a
polynomial expression with a finite number of terms, greater than
two, we will have to introduce more free parameters to our model
which will necessarily be fixed by experimental data. These
additional terms will make the model less predictive, so it appears
that if we want a finite polynomial expression, it is most natural
to keep the AdS form of the field $X$. Indeed we will show that this
can be done consistently while finding the back-reaction on the
geometry.

The second option may have significant implications in the context
of linear confinement where a specific IR behavior for the metric
and dilaton have been discussed in \cite{hep-ph/0602229}. Finding a
back-reacted solution which gives this is an important goal and is
discussed more in section \ref{sec.conclusion}.

The third option provides an interesting prospect, by which the
space is truncated smoothly without the need of an IR boundary put
in by hand, though the interpretation of the additional field may
itself come from some distribution of branes in the IR. The benefit
of such a choice may be that the number of free parameters of the
model is reduced, by making the value of the IR cutoff depend
explicitly on the value of the quark bilinear condensate, as is
expected from QCD. The question however is, what function should one
choose for the $X$ field. We show in section \ref{ssec.br} that a
simple guess gives a consistent set of predictions.

\subsection{Model I}\label{sec.model1}

\begin{figure}
\begin{center}
\includegraphics[width=8cm,clip=true,keepaspectratio=true]{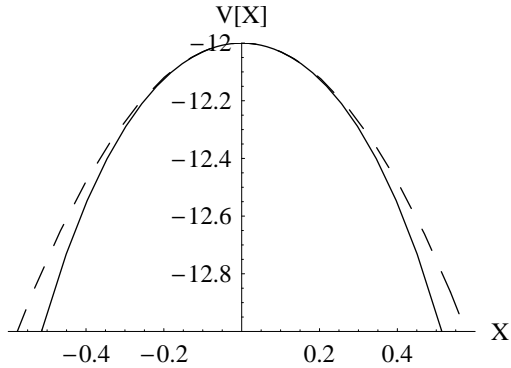}
\caption{The potential in the five-dimensional action for both the
non-back-reacted (dashed line) and back-reacted (full line)
theories.}\label{fig.plv}
\end{center}
\end{figure}

From Eqn(\ref{eq.8}) we can calculate the effect of the tachyonic
field given by Eqn(\ref{eq.X}) on the metric and find that the
back-reacted function $A$ is given by:
\begin{equation}
A(z)=\log(z)+\frac{m_q^2}{24}z^2+\frac{m_q
\sigma}{16}z^4+\frac{\sigma^2}{24}z^6
\end{equation}
Figure \ref{fig.plv} shows the potential in this back-reacted theory
compared to the potential without the back-reacton. The potential
goes roughly as $X^4$ for large $X$.
The inclusion of the back-reaction now means that the mass of the
vector mesons will be a function of the quark mass and quark
bilinear condensate. The question is, how much of an effect this
will be. One expects in QCD that the mass of the vector mesons
should be determined from the value of $\Lambda_{QCD}$ and should
not be affected significantly by the quark parameters.

Just as in \cite{hep-ph/0501128} we choose to fix the four free
parameters in this theory using $N_c,f_\pi,M_\pi$ and $M_\rho$. A
simple check shows that, as expected, the addition of the
back-reaction does not affect the proof that the
Gell-Mann-Oakes-Renner relation holds in this model. This
immediately allows us to fix the quarks bilinear condensate as a
function of the quark mass, $f_\pi$ and $M_\pi$ (errors will be of
the order $m_q^2$ and therefore negligible in any reasonable model).
The five-dimensional coupling is fixed from the vector current
two-point function to be
\begin{equation}
g_5^2=\frac{12\pi^2}{N_c}\ ,
\end{equation}
and for the phenomenological model we pick $N_c=3$. This leaves us
with two free parameters, $m_q$ and $z_{IR}$, the IR cutoff.

To fix $z_{IR}$ we use the position of the pole in the vector
current correlator, provided by the eigenvalue of its dual
five-dimensional field. The equation of motion for an arbitrary
component of the vector field, $\psi(z)$, is given by
Eqn(\ref{eq.vect}).
\begin{equation}\label{eq.vect}
12 e^{\frac{1}{24} \left(2 \sigma ^2 z^6+3 \sigma  m_q z^4+2 m_q^2
z^2\right)} z \psi(z) m_{\rho }^2-\left(3 \sigma ^2 z^6+3 \sigma m_q
z^4+m_q^2 z^2+12\right) \psi'(z)+12
   z \psi''(z)=0\ ,
\end{equation}
where we have the constraint that
\begin{equation}
\sigma=\frac{M_\pi^2f_\pi^2}{2m_q}\ .
\end{equation}
The boundary conditions for the field $\psi$ are given by a
vanishing derivative at the IR-brane and a vanishing value in the UV
(at $z\rightarrow 0$). In order to calculate the $\rho$-meson decay
constant, we must be careful of the normalization, however, the
normalization will not affect the position of the zeros of the
function in the UV and we are therefore free to choose
$\psi(z_{IR})=1$. Having fixed the IR conditions of $\psi$
completely, we can vary $z_{IR}$ and $m_q$ to find which values give
the correct $z\rightarrow 0$ behavior. By plotting the value of
$\psi(\epsilon)$ as a function of the two free parameters on a scale
such that all we can see is the switch from positive to negative
$\psi(\epsilon)$, it is clear that this is almost independent of the
quark mass, for a reasonable range of values.
\begin{figure}
\begin{center}
\includegraphics[width=8cm,clip=true,keepaspectratio=true]{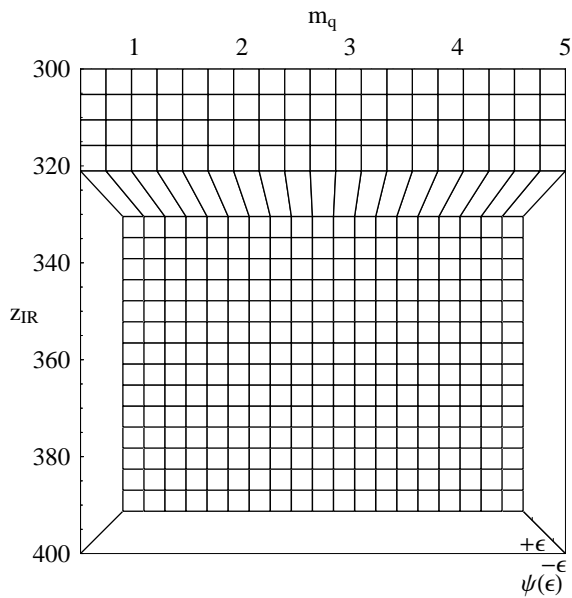}
\caption{$\psi(\epsilon)$ as a function of $z_{IR}$ and $m_q$,
showing the contour which corresponds to the physical boundary value
for this field.}\label{fig.pl1}
\end{center}
\end{figure}
This result is as expected and tells us that the vector spectrum is
determined largely through $\Lambda_{QCD}$ and not through $m_q$.

In order to fix the two parameters completely we now turn to the
pion decay constant, calculated from the axial vector equation of
motion with the pole in the propagator set to zero. Note that the
axial vector equation of motion depends on the quark mass and
condensate even neglecting the back reaction. The pion decay
constant is given by:
\begin{equation}
f_\pi^2=\left.-\frac{1}{g_5^2}\frac{\partial_z
A(q=0,z)}{z}\right|_{z=\epsilon}\ ,
\end{equation}
Where the solution of the axial field $A$ is the solution of
\begin{equation}
48 \pi ^2 z A(z) \left(\sigma  z^2+m_q\right)^2+\left(3 \sigma ^2
z^6+3 \sigma  m_q z^4+m_q^2 z^2+12\right) A'(z)-12 z A''(z)=0\ .
\end{equation}
This time we plot the value of $f_{\pi,mod}-f_{\pi,exp}$ as a
function of the two free parameters, where $f_{\pi,mod}$ corresponds
to the value calculated in this model and $f_{\pi,exp}$ corresponds
to the experimental value. By combining this plot with the plot in
figure \ref{fig.pl1} we can fix the values of the two free
parameters precisely. These are found to be $m_q=2.4 MeV$ and
$\sigma=(323.5 MeV)^3$.
\begin{figure}[!h]
\begin{center}
\includegraphics[width=8cm,clip=true,keepaspectratio=true]{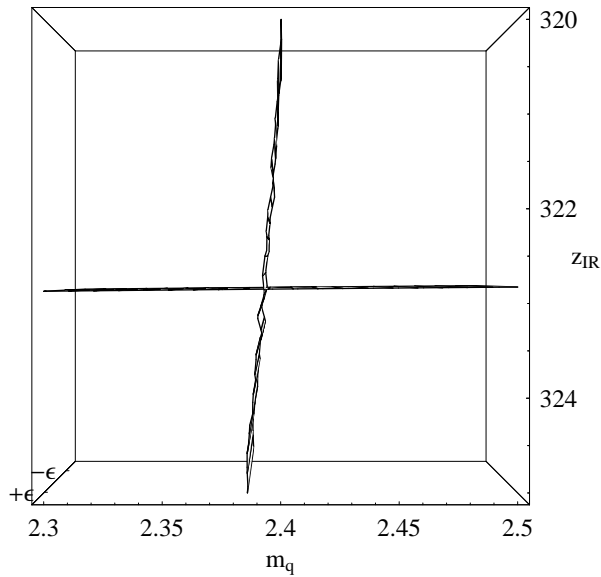}
\caption{Intersection of the physical contours for both $f_\pi$ and
$\psi(\epsilon)$ as a function of the two free parameters $z_{IR}$
and $m_q$. The point of intersection fixes the values to be used to
calculate the observable predictions.}\label{fig.plall}
\end{center}
\end{figure}
We fix the remaining free parameters of the theory from the
intercept values of the two graphs and are now ready to calculate
the remaining observable quantities, precisely as in
\cite{hep-ph/0501128}. The results of this analysis are given in
table \ref{tab.results}.
\begin{table}
\centering
\begin{tabular}{||l|l|l|l|l||}
  \hline
  Observables & Measured Value & Non-back-reacted  & Model I results& Model II results\\
  &  (MeV)& results (MeV (\% error)) &(MeV (\% error)) &(MeV (\% error))\\
  \hline
  \hline
  $m_\pi$ & $139.6\pm0.0004$ & $139.6*$& $139.6*$& $139.6*$\\
  $f_\pi$ & $92.4\pm 0.35$ & $92.4*$ & $92.4*$& $92.4*$\\
  $m_\rho$ & $775.8\pm0.5$ & $775.8*$& $775.8*$& $769 (0.9)$\\
  $\sqrt{F_\rho}$ & $345\pm 8$& $329 (4.6)$ & $334 (3.2)$& $325.5 (5.6)$\\
  $m_{a_1}$ & $1230\pm 40$& $1363 (10.8)$ & $1348 (9.5)$& $1403 (14)$\\
  $\sqrt{F_{a_1}}$ & $433\pm 13$ & $486 (12.2)$ & $481 (11)$& $514 (18.8)$\\
  $g_{\rho\pi\pi}$ & $6.03\pm 0.07$ &$4.44 (26)$ & $4.46 (26)$& $4.49 (25)$\\
  \hline
  \hline
  rms error & - & 15\% & 15\%& 16\%\\
  \hline
  \hline
\end{tabular}
\caption{Results of the non-back-reacted, non-singular back-reacted
(Model I) and singular back-reacted (Model II) geometries. The
percentage error is taken from the central value of the experimental
measurement. $*$ indicates that this observable was used to fit the
data.}\label{tab.results}
\end{table}

We see that the effect of the back-reaction in this case is very
small. We have shown that we can construct a phenomenologically
consistent model, including back-reaction and obtain results in very
close agreement with those without the back-reaction. The reason for
this is simple. The effect of the back-reaction on the geometry will
only become significant for energies of the order of
$\sigma^{\frac{1}{3}}$. With the inclusion of the IR-brane, the
geometry is cutoff just below the region where these effects would
become significant and so the effects of the quark dynamics are
shown to be negligible.

\subsection{Model II}\label{ssec.br}
As stated in the previous section, it would be appealing to remove
$z_{IR}$ as a free parameter and instead to generate it dynamically
via a condensate. We examine one possible function with which we can
do this and show that the results are promising.

Motivated by supergravity solutions we make a simple choice for $X$,
of the form
\begin{eqnarray}
X&=&\left(\frac{m_q z}{2}+\frac{1}{4} \log[\frac{1+\sigma
z^3}{1-\sigma z^3}]\right){\mathbb
I}_{2\times 2}\ , \nonumber\\
\lim_{z\rightarrow 0}&\sim&\ \left(\frac{m_qz}{2}+\frac{\sigma
z^3}{2}\right){\mathbb I}_{2\times 2}\ ,
\end{eqnarray}
where the limiting behavior is for small $z$. The space is therefore
cutoff at $z_{IR}=\frac{1}{\sigma^{\frac{1}{3}}}$.

Again we use the Gell-Mann-Oakes-Renner relation to constrain the
relationship between $m_q$ and $\sigma$ leaving us with one free
parameter which we can fix using the pion decay constant. The values
are found to be $m_q=2.53$ $MeV$ and $\sigma=(320.3 MeV)^3$.

We perform the same analysis using this model as we did in section
\ref{sec.model1}. The results of this analysis are given in table
\ref{tab.results}. This singular back-reacted geometry gives good
results and the important point is that the singular behavior does
not destroy the original phenomenology. There are clearly many other
sensible choices for $X$ which are both singular and give the
correct UV behavior, however we show that with a sensible choice,
such a model is phenomenologically consistent.

\begin{figure}
\begin{center}
\includegraphics[width=10cm,clip=true,keepaspectratio=true]{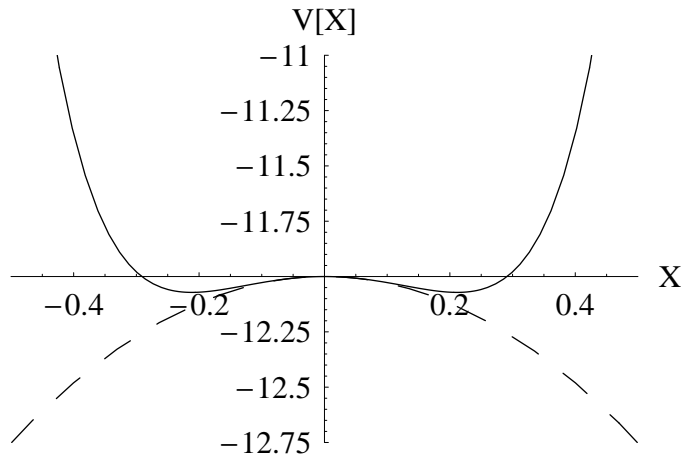}
\caption{The potential in the five-dimensional action for both the
non-back-reacted (dashed line) and back-reacted (full line), Model
II, theories.}\label{fig.plv2}
\end{center}
\end{figure}

We can study the potential in the case of this singular back-reacted
geometry and compare it to the original potential in the
non-back-reacted case. This is shown in figure \ref{fig.plv2}. We
see that while the original potential is unbounded, the potential
which classically gives us the singular field, $X$, gives a bounded
potential with a minimum. The position of the minimum is a function
of $m_q$ and $\sigma$ but its existence is not dependent on these
values.
Though we treat the five-dimensional theory classically, from a
quantum mechanical perspective, the potential in this singular,
back-reacted geometry would have important consequences.

\section{Discussion and Conclusion}\label{sec.conclusion}
In this paper we consider the back-reaction of the scalar on the
metric in the holographic QCD model. In the hard wall model, the
meson spectrum is generated by the IR brane. For pure AdS without
the hard wall, the five-dimensional model is scale invariant and all
the mesons are massless. Breaking the scale invariance is an
essential mechanism in order to generate the meson spectrum. The
mass gap $\Lambda_{QCD}$ is dual to the IR cut-off in this model.
After considering the back-reaction, we find that the model still
successfully describes the spectrum and decay constants of ground
state mesons.

As for the resonances, in \cite{hep-ph/0602229} it was shown that
one can get linear confinement by including a quadratic dilaton
$\Phi \sim z^2$ in the pure AdS background (see also
\cite{hep-ph/0609112} for a phenomenological treatment including a
UV cutoff). The peculiar $z^2$ dependence of the dilaton is far from
clear. If one considers the deformed AdS metric without introducing
a dilaton:
\[
ds^2={f(z) \over z^2} (dx_{\mu}dx^{\mu}-dz^2),
\]
it can be shown that with $f(z) \sim (\cosh(az^2) + b
\sinh(az^2))^4$ and $a\neq0$, one breaks scale invariance and gets
linear confinement, even without the presence of the IR brane.
However, the deformed AdS metric given by the back-reaction of the
quark condensate is not compatible with this. Linear confinement can
not be realized in this phenomenological model. From the
non-polynomial form of $f(z)$, the first non-trivial term is
quadratic in $z^2$. This suggests that a non-local dimension two
condensate should be formed. In QCD, it has been investigated
\cite{Gubarev:2000eu} that the non-zero value for the minimum of the
vector potential condensate $A^2$, which is defined to be $\int d^3x
\vec{A}^2 (x)$ is possible. This dimension two condensate can be
shown to be gauge invariant but non-local. Therefore, its effect
should be relevant to linear confinement.

\section*{Acknowledgements}

This work was supported in part by the key projects of Chinese
Academy of Sciences, the National Science Foundation of China (NSFC)
under the grant 10475105, 10491306. This research was also supported
in part by the National Science Foundation under Grant No.
PHY99-07949.

\end{document}